\documentclass[%
reprint,
 amsmath,amssymb,
 aps,
pra,
]{revtex4-2}

\usepackage{graphicx}
\usepackage{dcolumn}
\usepackage{bm}
\usepackage{siunitx}

\begin{document}

\preprint{APS/123-QED}

\title{A Diffusive Model for Radio-Frequency Knock-Out Slow Extraction}

\author{P.~A.~Arrutia Sota}\thanks{pablo.arrutia@cern.ch}
\author{E.~C.~Cortés García}
\author{V.~A.~Sansipersico}
\affiliation{%
CERN, Geneva, Switzerland
}%

\date{\today}

\begin{abstract}
Slow resonant extraction from synchrotrons via radio-frequency knock-out is a well-established technique to deliver charged particle beams for various applications.\
In this contribution, we present explicit analytical expressions for calculating the number of particles slowly extracted over time—commonly referred to as spills.\
The proposed formulation enables the semi-analytical determination of an amplitude modulation curve to be applied to the radio-frequency exciter, which flattens the spill macrostructure, a feature of high relevance to all users requiring uniform beam delivery.
\end{abstract}

\maketitle




\textbf{I. Introduction.} Charged particle beams prepared in synchrotrons are often required to be delivered over extended timescales, typically from a few hundred to several thousand milliseconds.\
To achieve this slow extraction—much longer than the revolution period and independent of the bunch structure—the resonant extraction technique is employed.\
Among the possible resonance orders, the third-order resonance has become the standard working point~\cite{KobayashiHamiltonian}, widely adopted for applications such as electronics irradiation~\cite{söderström:ipac2025-tupb006}, cancer therapy~\cite{medsyncNoda}, and fixed-target experiments.\
Several variations of this method exist; one particularly effective approach for enhanced control and quality of the extracted particle counts, so called spills, is the radio-frequency knock-out (RF-KO) technique~\cite{TOMIZAWA1993}.

In RF-KO extraction, the beam optics are kept constant while the betatron motion of particles is deliberately driven by an external excitation.\ 
This induces a controlled emittance growth, a process extensively studied over the past decades to optimize both the macro- and micro-spill structures (see e.g.~\cite{NodaDualFM, PNiedermayer2024,ArrutiaSota:2024vlt}).\
Owing to the complexity of single-particle dynamics near resonance, comprehensive analytical models remain out of reach.\ 
Thus, optimization relies on particle tracking and/or dedicated machine development campaigns for fine tuning.\

In this contribution, we adopt a simplified diffusive-model approach to RF-KO extraction~\cite{vanderMeer1978,FURUKAWA2004} and derive analytical expressions to describe the spill.
These are applied to reconstruct the spill profile and to compute an amplitude modulation curve for the excitation, aiming to achieve a flat spill. The results are validated through experimental campaigns conducted at the CERN Proton Synchrotron (PS).\
The paper is organized as follows: Section II summarizes the established effective model describing the beam dynamics during RF-KO extraction.\ 
Section III presents the diffusive model and its analytical formulation, followed by a comparison between the two approaches.\ 
Section IV discusses the amplitude modulation required to obtain a constant spill flux.\
Section V presents the experimental data validating the results, and Section VI concludes with a summary of the main findings.

\textbf{II. Map-based model.} Numerical simulations of the RF-KO process are typically performed by employing map-based models, i.e. tracking codes.
In this contribution, we will use a simple `Hénon-like' map as our baseline model,
which will later be used to benchmark the proposed diffusive model. 
The map updates the particle's normalized transverse offset $X$ and momentum $P$
every turn $n$ according to,
\begin{equation}
\begin{bmatrix}
X_{n+1} \\
P_{n+1}
\end{bmatrix}
=
R(\mu)
\begin{bmatrix}
X_n \\
P_n + S X^2_n + k(t)
\end{bmatrix}
\end{equation}
where $S$ is the resonance driving term, $k(t)$ is the time-dependent exciter kick, and $R(\mu)$ is the 2-dimensional rotation matrix by an angle $\mu$,
\begin{equation}
  \mu = 2\pi\left(m + \frac{1}{3} + \delta Q\right), \ \ m \in \mathbb{Z},
\end{equation}
with $\delta Q$ being the tune distance to the third-integer tune, which is taken to be the same for all particles.\
For the CERN PS, $m=6$ and $S=77 \text{m}^{-\frac{1}{2}}$. 

\begin{figure}[ht]
  \centering
  \includegraphics[width=0.35\textwidth]{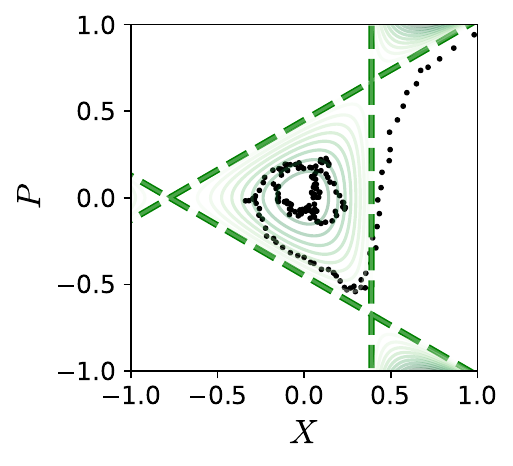}
  \caption{Hamiltonian contours (greens) and time evolution of a single initial condition under RFKO kicks (black), plotted every three turns. }
  \label{fig:phase_space}
\end{figure}

When $k(t) = 0$, the dynamics may be smoothed and averaged over three turns to obtain the so-called `Kobayashi' Hamiltonian~\cite{KobayashiHamiltonian},
\begin{equation}
  H = 3\pi \delta Q \left(P^2 + X^2\right) + \frac{S}{4} \left(3XP^2 - X^3\right),
\end{equation}
which defines the phase-space geometry shown in Fig.~\ref{fig:phase_space}. 
As it can be seen, a triangular separatrix divides the space into a stable and an unstable region.
The exciter kicks may then be seen as perturbations, 
transporting particles from the stable to the unstable region to deliver the desired extraction flux.

\textbf{III. Diffusive model.} Motivated by the phase-space geometry and dynamics described above, alternatively we propose to model the extraction process with a diffusion equation in cylindrical coordinates~\cite{wiedemann_particle_2015}:
\begin{equation}
\partial_t u = \frac{1}{r}\partial_r(r D\partial_r u),
\label{eq:diffusion_equation}
\end{equation}
where $u(r,t)$ represents the beam distribution, with $r = \sqrt{X^2 + P^2}$. The angular dynamics have been ignored, even though the geometry in Fig.~\ref{fig:phase_space} is not perfectly rotational-symmetric (especially near the separatrix boundary).

In general, we will allow the diffusion coefficient $D$ to be time-dependent and define it by the relationship \mbox{$2D \cdot t = \langle r^2 \rangle$}. For example, if the exciter kicks are drawn independently
from a zero-mean Gaussian distribution with standard deviation $k$, we obtain $D = k^2/4$.
In practice, exciters will have limited bandwidth, but the formalism will still remain useful, as we will demonstrate in later sections. 

To represent the impact of the separatrix, we impose an absorbing boundary condition $u(r=R, t) = 0$
at the outer radius $R$, computed by equating the area inside the stable region (triangle) to the area of a circle of radius $R$,
i.e. $R = \left(48 \sqrt{3} \pi\right)^{1/2} \frac{\delta Q}{S}$.

\textit{A.Time-dependent solution.}~A general solution to Eq.~\ref{eq:diffusion_equation} can be obtained by the method of separation of variables~\cite{olver_introduction_2014}:
\begin{equation}
u(r,t) = \sum_{i=1}^{\infty} a_i\,\exp\left[-\lambda_i^2\,\Theta(t)\right] J_0\left(\lambda_i r\right),
\end{equation}
where $\Theta(t) \equiv \int_0^t D(\tau)\,d\tau$, $\lambda_i=\zeta_i/R$, $J_0$ is the Bessel function of the first kind of order zero, 
and $\zeta_i$ are the positive zeros of $J_0$ (i.e., $J_0(\zeta_i) = 0$). The coefficients $a_i$ are obtained by projecting the initial condition $u_0(r)$ onto the basis functions:
\begin{equation}
a_i = \frac{2}{R^2 [J_1(\zeta_i)]^2}\int_0^R u_0(r) J_0\left(\lambda_i r\right) r dr,
\end{equation}
where $J_1$ is the Bessel function of the first kind of order one and $\int_0^R u_0(r) r dr = N$, the initial number of particles.

A typical initial condition is the truncated Gaussian distribution:
\begin{align}
u_0(r) = \begin{cases}
    \frac{1}{A}\exp\left[-\frac{1}{2}\left(\frac{r}{R\sigma}\right)^2\right] \quad &\text{for } r < R,\\
    0 \quad &\text{for } r \geq R,
\end{cases}
\end{align}
where $\sigma$ is the standard deviation of the corresponding non-truncated Gaussian in units of $R$ and
\begin{equation}
A = 2\pi\frac{R^{2}\sigma^{2}}{N}\left[1-\exp\left(-\frac{1}{2\sigma^{2}}\right)\right].
\end{equation}
Typically, $\sigma \ll R$ such that the particle density near the boundary is small.


\textit{B. Slow-extracted flux.}
Within this diffusive model, the slow-extracted flux is given by the outward flux at the boundary:
\begin{equation}
\Phi(t) = -2\pi R D(t) \frac{\partial u}{\partial r}\Big|_{r=R} = 2\pi D(t) \sum_{i=1}^\infty b_i\,e^{-\lambda_i^2\Theta(t)},
\end{equation}
with $b_i = a_i\,\zeta_i\,J_1(\zeta_i)$.

\begin{figure}[ht]
  \centering
  \includegraphics[width=0.5\textwidth]{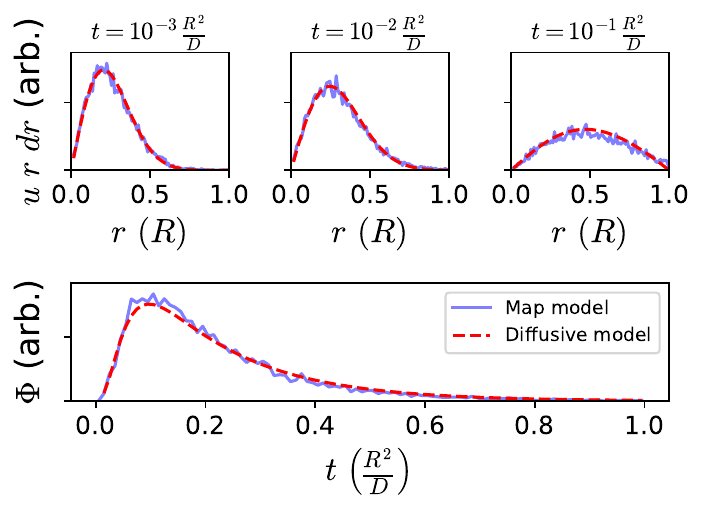}
  \caption{Comparison between map-based model (blue) and diffusive model (red) showing
  radial distribution evolution at different times (top row) and total flux evolution (bottom row).}
  \label{fig:spill_distribution}
\end{figure}

Figure~\ref{fig:spill_distribution} shows the outcome of a numerical experiment,
where the diffusive model is compared to the map-based model in the case of constant diffusion coefficient $D_0$,
i.e. $\Theta(t) = D_0 \cdot t$. Good agreement is observed both when comparing the radial distributions and the extracted flux. The flux is strongly time-dependent, which is typically undesirable for experimental users.

\textbf{IV. Obtaining Constant Flux.}
We now aim to achieve a constant extracted flux $\Phi_0$ by appropriately varying $D(t)$.
Therefore, we require:
\begin{equation}
D(t) = \frac{d\Theta}{dt} = \frac{\Phi_0}{2\pi\,S(\Theta)}, \quad S(\Theta) \equiv \sum_{i=1}^{\infty} b_i\,e^{-\lambda_i^2\Theta},
\label{eq:D_equation}
\end{equation}
which is an autonomous ordinary differential equation for $\Theta$, yielding
\begin{equation}
t(\Theta) = \frac{2\pi}{\Phi_0}\sum_{n=1}^{\infty}\frac{b_i}{\lambda_i^2}\left(1-e^{-\lambda_i^2\Theta}\right).
\end{equation}

The time-dependent diffusion coefficient is then obtained by numerically inverting $t \mapsto \Theta(t)$ within
the interval $[0, t^*]$, with $t^* = N/\Phi_0$, i.e.~the time it takes to extract all particles $N$.\
Then, the results are substituted back into Eq.~\ref{eq:D_equation} to obtain the time-dependent diffusion coefficient. 

\begin{figure}[hbt]
  \centering
  \includegraphics[width=0.5\textwidth]{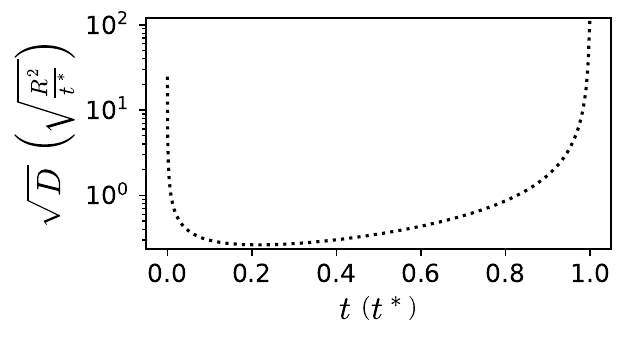}
  \caption{Square root of the time-dependent diffusion coefficient $D(t)$ for an initial distribution with $\sigma=1/5$.}
  \label{fig:d_of_t}
\end{figure}

Figure~\ref{fig:d_of_t} shows the output of this procedure for an initial distribution with $\sigma=1/5$
where $\sqrt{D(t)}$ has been plotted, as it relates more directly to the kick amplitude of the exciter.
It can be seen that the curve exhibits a `bathtub' shape,
qualitatively consistent with the empirical solutions found by other RF-KO practitioners~\cite{vanderMeer1978, FURUKAWA2004}.

\begin{figure}[ht]
  \centering
  \includegraphics[width=0.5\textwidth]{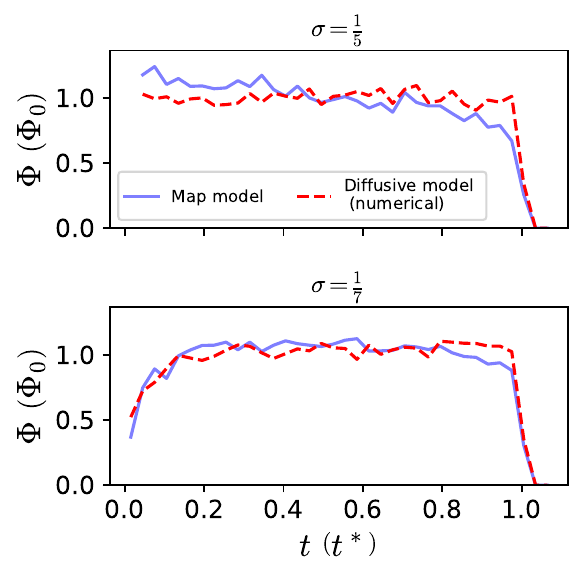}
  \caption{Extracted flux $\Phi(t)$ for the map-based model (blue) and numerically integrated diffusive model (red)
  with time-dependent diffusion coefficient as specified by Eq.~\ref{eq:D_equation} to maintain
  $\Phi (t) = \Phi_0$.}
  \label{fig:constant_flux}
\end{figure}

Figure~\ref{fig:constant_flux} shows the outcome of a numerical experiment
where $k(t)$ has been varied according to Eq.~\ref{eq:D_equation} to achieve $\Phi_0 = N/t^*$ with $t^* = 10^5$ turns. Both the map model and the numerically integrated diffusive model yield spills with markedly improved flatness. Some differences remain, arising from the simplifying assumptions in Section III (e.g., perfect rotational symmetry). Nonetheless, the spill flatness achieved by the map model is adequate for most practical applications.\ Otherwise, the theoretical estimate of $k(t)$ may be used as an initial input for a feedback/feedforward routine.

\textit{A. Asymptotic Behavior and Physical Constraints.} For late times, the $i=1$ mode dominates in Eq.~\ref{eq:D_equation}, giving $S(\Theta) \rightarrow b_1 e^{-\lambda_1^2\Theta}$. 
Then, for a given flux $\Phi_0$, the system approaches a finite horizon time:
\begin{equation}
t_\ast = \frac{2\pi}{\Phi_0}\sum_{i=1}^{\infty}\frac{b_i}{\lambda_i^2} = \frac{N}{\Phi_0}.
\end{equation}
and, as $t \rightarrow t_\ast$, the diffusion coefficient diverges as:
\begin{equation}
D(t) \rightarrow \frac{1}{\lambda_1^2\left(t_*-t\right)}.
\label{eq:D_asymptotic}
\end{equation}

Unsurprisingly, the calculation verifies that one cannot extract more particles than the initial number $N$. 
More importantly, this result reveals that a truly constant flux over the whole beam is not achievable for an exciter system with finite power as the required $D$ becomes unbounded towards the end of the extraction. 

\textbf{V. Test in the CERN PS.} A measurement test was performed in the CERN PS to validate the model.\
Details of the machine can be found elsewhere e.g.~\cite{VanGoethem:2025hib}.
In this case, the exciter signal is given by a series of amplitude-modulated chirps,
\begin{equation}
k(t) = k_0(t) \sin\left[2\pi \left(q_0 + \frac{\Delta q}{T} \tau\right) \frac{t}{T_0} + \phi_0\right], \quad \tau = t\!\!\!\!\mod\!T,
\end{equation}
where $k_0(t)$ is the amplitude of the chirp, $q_0$ is the initial exciter tune, $\Delta q$ is the tune sweep,
$T$ is the period of the chirp, $T_0$ is the revolution period, and $\phi_0$ is the initial phase.

Within a single chirp, the kicks received by a particle from one turn to another are strongly correlated.\
However, the amplitude growths from consecutive chirps are uncorrelated, and we may therefore expect diffusive behavior over timescales much longer than the chirp period.\ 
In fact, the diffusion coefficient can be computed analytically~\cite{jones_comparison_1959}, yielding $D = k_0^2/(16\Delta q T_0)$.\
Interestingly, the result is independent of $T$, since a single chirp 
with longer $T$ leads to more growth, but a chirp with shorter $T$ is repeated at a higher frequency.\

\begin{figure}
    \centering
    \includegraphics[width=0.4\textwidth]{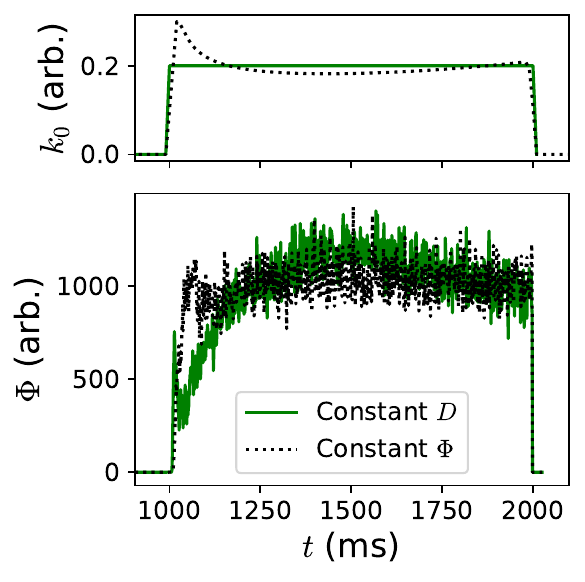}
    \caption{Experimental measurement of the extracted spill at the CERN Proton Synchrotron using a Pb$^{54+}$ beam with $E_{\text{kin}} =$\SI{1}{\giga e \volt / nucleon}.\
    Results are shown for both constant and time-dependent exciter gains.\
    Each curve represents the average over 10 acquisitions of counts recorded by a scintillator screen. The integration time is one millisecond.}
    \label{fig:experiment}
\end{figure}

Figure~\ref{fig:experiment} shows the outcome of a beam test performed at the CERN PS with a Pb$^{54+}$ beam with $E_{\text{kin}} =$\SI{1}{\giga e \volt / nucleon}.\ 
The relevant beam parameters and machine optics were fully characterized using standard methods.
The normalized horizontal and vertical RMS emittances were determined to be $\varepsilon_{x,n} = (6.47\pm 0.10) $\SI{}{\micro\meter} and $\varepsilon_{y,n} = (2.42\pm 0.01)$\SI{}{\micro\meter}, respectively. The total injected beam intensity was $I = 1.64\times10^{9}$ ions with an RMS relative momentum spread of $\sigma_p = 0.75 \times 10^{-3}$.\ 
The machine was operated with transverse tunes close to the horizontal third-order resonance, at $(Q_x, Q_y)=(6.321, 6.227)$.\
The corrected operational chromaticities were determined to be $(\xi_x, \xi_y) = (-0.01,-0.12)$.



As illustrated in Fig.~\ref{fig:experiment}, the flux variations throughout the spill were corrected by replacing the nominal gain curve (constant $D$) with the analytically estimated time-dependent gain curve (constant $\phi$).\ 
It can be seen that the new configuration successfully improves the spill structure, while keeping the duration and number of extracted particles unaffected.

\textbf{VI. Conclusion and summary.}
A diffusive model has been developed for RF-KO slow extraction. 
The model can be used to make analytical predictions of the 
beam distribution and extracted flux as a function of time, 
as well as to estimate the time-dependent amplitude modulation
required to deliver a constant-flux spill. The calculations have
been benchmarked with a numerical map model and with measurements
in the CERN PS, demonstrating good agreement.
The model paves the way for faster and more efficient optimization of global spill control, as well as providing an alternative framework to conceptualize RFKO slow-extraction. 

\textbf{Acknowledgments.} We warmly thank C. Bracco and M. Fraser for proofreading the manuscript and for their helpful suggestions. We also extend our gratitude to the operators and the Beam Transfer Physics colleagues for the valuable discussions, especially M. Remta. ECCG acknowledges funding from the HEARTS project, which is funded by the EU under GA No 101082402.

\bibliography{main}

\end{document}